\documentclass[reprint,aps,pra,superscriptaddress,twocolumn,showpacs]{revtex4}
\usepackage{epsfig}
\usepackage{graphics}
\usepackage{amsmath}
\usepackage{subfigure}
\usepackage{color}
\usepackage{float}
\usepackage{soul}
\usepackage[usenames, dvipsnames]{xcolor}

\begin{document}

\title{Stable multiple vortices in collisionally inhomogeneous attractive
Bose-Einstein condensates}
\author{J. B. Sudharsan}
\author{R. Radha}
\email{radha_ramaswamy@gcwk.ac.in}
\affiliation{Centre for Nonlinear Science (CeNSc), Post-Graduate and Research Department
of Physics, Government College for Women (Autonomous), Kumbakonam 612001,
India.}
\author{H. Fabrelli and A. Gammal}
\affiliation{Instituto de F\'isica, Universidade de S\~ao Paulo, 05508-090, S\~ao Paulo,
Brazil.}
\author{Boris A. Malomed}
\affiliation{Department of Physical Electronics, School of Electrical Engineering, Tel
Aviv University, Tel Aviv 69978, Israel.}

\begin{abstract}
We study stability of solitary vortices in the two-dimensional trapped
Bose-Einstein condensate (BEC) with a spatially localized region of
self-attraction. Solving the respective Bogoliubov-de Gennes equations and
running direct simulations of the underlying Gross-Pitaevskii equation
reveals that vortices with topological charge up to $S=6$ (at least) are
stable above a critical value of the chemical potential (i.e., below a
critical number of atoms, which sharply increases with $S$). The largest
nonlinearity-localization radius admitting the stabilization of the
higher-order vortices is estimated analytically and accurately identified in
a numerical form. To the best of our knowledge, this is the first example of
a setting which gives rise to \emph{stable} higher-order vortices, $S>1$, in
a trapped self-attractive BEC. The same setting may be realized in nonlinear
optics too. \newline
\end{abstract}

\pacs{03.75.Lm; 05.45.Yv; 42.65.Tg}
\maketitle

\textit{Introduction.} The creation of atomic Bose-Einstein condensates
(BECs) \cite{carlwiemanncornell}, and subsequent identification of nonlinear
excitations in them, such as bright \cite{bright,bright-reviews} and dark
solitons \cite{dark,dark-review} in quasi-one-dimensional ``cigar-shaped"
settings, has triggered a great deal of interest in the investigation of
matter-wave dynamics in the BEC, which is very accurately modelled by the
Gross-Pitaevskii equation (GPE) \cite{pitaevskii}-\cite{gammalPM}. In its
general form, it is tantamount to the three-dimensional (3D) nonlinear Schr%
\"{o}dinger equation which includes a trapping potential and a cubic term
accounting for the mean-field nonlinearity.

The discovery of robust coherent nonlinear excitations, in the form of
\textit{dromions} \cite{dromions} (localized 2D patterns produced by
overlapping of 1D ghost solitons) and \textit{lumps} (weakly localized 2D
solitons) \cite{lumps} in other 2D models has prompted looking for similar
nonlinear excitations in BEC as well, following the investigation of more
straightforward nonlinear modes -- in particular, vortices \cite{vortices}
and Faraday waves \cite{FWs} -- in effectively 2D ``pancake-shaped" BECs.
Faraday waves are undulating nonlinear excitations generated by
time-periodic shaking of the trapping potential, while vortices are created
by stirring the condensate with the help of properly designed laser beams,
by coherent transfer of orbital angular momentum to the condensate by the
two-photon stimulated Raman process \cite{vortices,Fetter}, or by imprinting
an appropriate phase pattern onto a trapped condensate \cite{magtrap}, see
recent survey \cite{PNAS}. While in self-repulsive condensates, simple
vortices (with topological charge $1$) are normally stable, their stability
in attractive BEC is a challenging problem, as the self-attraction gives
rise to both collapse and azimuthal instability of solitary vortices, thus
easily destroying them \cite{Dalfovo}-\cite{review2}. Vortex solitons tend
to be unstable even in media where the collapse does not occur, such as
those featuring the quadratic nonlinearity \cite{Dima-Dima}.

The identification and experimental demonstration of more complex vortical
structures, such as vortex dipoles \cite{dipoles,dipoles-exper} and
quadrupoles \cite{quadrupoles,Paderborn}, was confined to BEC with the
self-repulsive nonlinearity. The most natural setting for hosting vortices
is provided by a pancake-shaped\ axially symmetric BEC, which is strongly
confined in one direction ($z$) and weakly confined in the transverse plane.
Although stable vortices with topological charge $S=1$ have been predicted
in the 2D self-attractive BEC with an in-plane trapping potential \cite%
{attractiveBEC-earlier}-\cite{brtka}, all multiple vortices with $S\geq 2$
were found to be unstable in the same model. This occurs by virtue of the fact that multiple vortices are vulnerable to instability against
splitting into unitary ones (in repulsive media too \cite%
{Neu,splitting,Fetter}).

Stabilization of multiple vortices is a fundamentally interesting problem
\cite{multiple}. The present work aims to predict stable vortices with $%
S\geq 2$ in a condensate with \emph{spatially localized} attractive
nonlinearity, which, without the help of the trapping potential, may support
stable fundamental solitons (with $S=0$), but not vortices \cite{HS,review2}%
. To the best of our knowledge, the setting elaborated in the present work
is the first system which gives rise to stable higher-order vortices in the
trapped self-attractive BEC. The model may also find its realization in
nonlinear optics.

\textit{The model and basic results.} In the mean-field approximation, the
self-attractive BEC, weakly trapped in the radial direction and strongly
confined along the $z$ axis, is governed by the 3D GPE for the single-atom
wave function \cite{pitaevskii, pethick}:
\begin{gather}
i\hbar \frac{\partial \Psi }{\partial T}=\left( -\frac{\hbar ^{2}}{2m}\nabla
_{\mathrm{3D}}^{2}+\frac{m}{2}\Omega _{z}^{2}[\Omega
_{r}^{2}(X^{2}+Y^{2})+Z^{2}]\right.  \notag \\
\left. +\frac{4\pi \hbar ^{2}a_{s}\mathcal{N}}{m}|\Psi |^{2}\right) \Psi ,
\label{eq1}
\end{gather}%
where $\nabla _{\mathrm{3D}}^{2}$ acts on coordinates $X,Y,Z$, $m$ is the
atomic mass, $\Omega _{r}$ and $\Omega _{z}$ are frequencies of
the radial and axial confinement (so that $\Omega _{r}$ is defined as the
relative frequency), $a_{s}$ is the $s$-wave scattering length, which is
negative for the attractive interatomic interactions, and $\mathcal{N}$ is
the number of atoms in the condensate, the norm of the wave function being $%
1 $. Factorizing the wave function by means of the usual substitution, $\Psi
\left(X,Y,Z,t\right) =\pi ^{-1/4}a_{z}^{-3/2}\exp \left( -i\Omega
_{z}t/2-Z^{2}/2a_{z}^{2}\right) \psi \left( x,y,t\right) $, where $a_{z}=%
\sqrt{\hbar /m\Omega _{z}}$ is the transverse-confinement size, one
integrates Eq. (\ref{eq1}) over $Z$ to derive the 2D form of the GPE \cite%
{perez-garcia,Luca,Canary} in terms of the scaled coordinates, $\left(
x,y\right) \equiv \left( X,Y\right) /a_{z}$, $t\equiv \Omega _{z}T$: $%
i\partial \psi /\partial t=\left[ -(1/2)\nabla _{\mathrm{2D}}^{2}+\left(
1/2\right) \Omega _{r}^{2}r^{2}+gN|\psi |^{2}\right] \psi $, where $\left(
r,\theta \right) $ are the polar coordinates in the $\left( x,y\right) $
plane, and $g\equiv 2a_{s}\sqrt{2\pi m\Omega _{z}/\hbar }$ is the effective
strength of the nonlinearity. The 2D wave function is also subject to the
unitary normalization condition, $\int \int |\psi (x,y,t)|^{2}dxdy=1$. We
here assume that the atomic density is not too high, therefore deviation of
the nonlinearity in the 2D equation from the cubic term \cite{Luca} may be
neglected.

It was predicted in various forms theoretically \cite{review2} and
demonstrated experimentally \cite{Japan} that the use of a Feshbach
resonance controlled by laser illumination \cite{Theis} makes it possible to
engineer spatially inhomogeneous nonlinearity \cite{review2}. Accordingly, the modified 2D GPE is rewritten as
\begin{equation}
i\frac{\partial \psi }{\partial t}=\left( -\frac{1}{2}\nabla _{\mathrm{2D}%
}^{2}+\frac{1}{2}r^{2}+g(r)N|\psi |^{2}\right) \psi ,  \label{cl2deq}
\end{equation}%
with the nonlinearity coefficient, $g$, made a function of the radial
coordinate, and $\Omega _{r}=1$ fixed by straightforward rescaling.

Stationary vortex solutions to Eq. (\ref{cl2deq}) are sought for as
\begin{equation}
\psi (r,\theta ,t)=R(r)\exp (iS\theta -i\mu t),  \label{radeq}
\end{equation}%
where $S$ is the integer vorticity and $\mu $ the chemical potential.
Inserting Eq. (\ref{radeq}) in Eq. (\ref{cl2deq}), we obtain an equation for
amplitude $R(r)$
\begin{equation}
2\mu R+R^{\prime \prime }+r^{-1}R^{\prime }-\left( S^{2}r^{-2}+r^{2}\right)
R-2gNR^{3}=0,  \label{gp_rad}
\end{equation}%
which can be solved numerically.

We choose a Gaussian spatial-modulation profile for the localized
self-attraction:
\begin{equation}
g(r)=-\exp \left( -b^{2}r^{2}/2\right) ,  \label{b}
\end{equation}%
where $b^{-1}$ determines the radius of the nonlinearity-bearing area, and a
free coefficient in front of the Gaussian was absorbed into $N$ in Eq. (\ref%
{cl2deq}), hence $N$ is proportional to the number of atoms, but is not
identical to it, unlike $\mathcal{N}$ in Eq. (\ref{eq1}). Thus, there remain
two free parameters in Eq. (\ref{cl2deq}), $b$ and $N$. While $N$
characterizes the strength of the nonlinearity, $1/b$ determines the the
radius of the nonlinearity-bearing region, in units of the trapping size
imposed by the in-plane harmonic-oscillator (HO) potential. The case of $%
b^{2}\ll 1$ amounts to the settings studied in works \cite%
{attractiveBEC-earlier,attractiveBEC}, which, as said above, admit stable
trapped states solely with $S=0$ and $1$. On the other hand, $b^{2}\gg 1$
implies that the system is almost linear, with a small nonlinear spot placed
at the center. In that case, the vortices are, essentially, the same modes
as their counterparts trapped in the HO potential in the framework of linear
quantum mechanics, hence they are stable. Thus, a nontrivial issue is to
find a minimum value, $b_{\min }$, at which the localization of the
nonlinearity leads to the stabilization of the vortices with $S\geq 2$. The
results reported below demonstrate that $b_{\min }\approx 1.0$ (in
particular, for $S=4$). This finding can be understood by noting that the
radial wave function of the linear HO, $R_{S}(r)=\left( \pi S!\right)
^{-1/2}r^{S}\exp \left( -r^{2}/2\right) $, yields average value $%
\left\langle r^{2}\right\rangle =1+S$. Then, the stabilization condition in
the framework of Eq. (\ref{cl2deq}) may be naturally defined as the
attenuation of the local strength of the nonlinearity by modulation profile (%
\ref{b}) by an order of magnitude (or more) at the location of the vortex, $%
r^{2}=$ $\left\langle r^{2}\right\rangle $ (this mechanism is different from the
centrifugal stabilization effect proposed in Ref. \cite{Dalfovo}, which was 
relevant for the uniform nonlinearity). Thus, the
stabilization is expected at\ $b\geq \sqrt{\left( 2/(1+S)\right) \ln 10}$.
In particular, for $S=4$ the estimate yields $b_{\min }\approx \allowbreak
0.96$, in reasonable agreement with the numerical findings.

In addition to BEC, the model based on Eqs. (\ref{cl2deq}) and (\ref{b}),
with $t$ replaced by the propagation distance, may be realized in optics as
a spatial-domain propagation equation in a bulk waveguide, with the
localized Kerr nonlinearity \cite{review2}, and the linear confining
potential representing transverse modulation of the refractive index in a
guiding channel. The localized nonlinearity may be provided by
inhomogeneously doping the host material with elements which induce local
enhancement of the nonlinearity via the two-photon resonance \cite{Kip}.

When the number of atoms is very small ($N\rightarrow 0$), the model
reduces, as said above, to the HO, with the respective eigenvalues, $\mu
_{\max }=1+S$. The contribution of the self-focusing nonlinearity to $\mu $
is negative, therefore $\mu _{\max }$ is the upper limit for values of the
chemical potential of trapped vortices.

The most essential results of the present work are summarized in Fig. \ref%
{mus} and Tables 1 and 2. In the figure, families of the vortices are
represented (including their stability, see details below) by dint of $\mu
(N)$ curves, for different values of $S$ at $b=1.5$. The results demonstrate
that the vortices are stable at $\mu _{\mathrm{\min }}<\mu <\mu _{\max }$,
where $\mu _{\min }$ is the boundary between solid and dashed segments in
Fig. \ref{mus}. At $\mu <\mu _{\min }$ the vortices are unstable, suffering
splitting and collapse (details are shown below). The figure also
demonstrates that values of $N$ corresponding to $\mu _{\min }$ increase
with $S$, which means that stable vortices with a higher topological charge
may hold much more atoms than their counterparts with $S=1$ and $2$. General
results for a range of relevant values of $b$ are collected in Tables 1 and
2.
\begin{figure}[h]
\centering
\subfigure{\includegraphics[height = 42.0 mm, width = 42.0
mm]{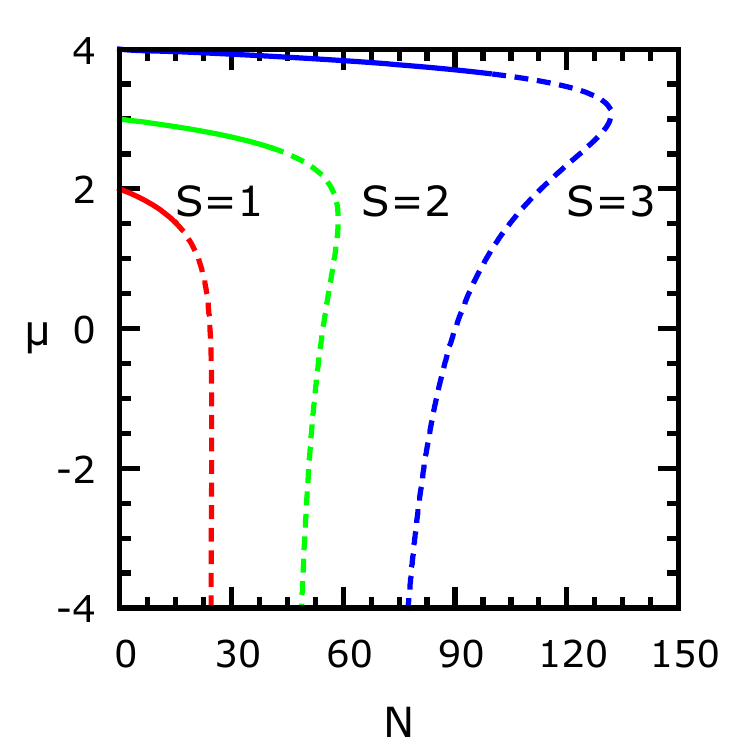}}
\subfigure{\includegraphics[height = 42.0 mm, width =
42.0 mm]{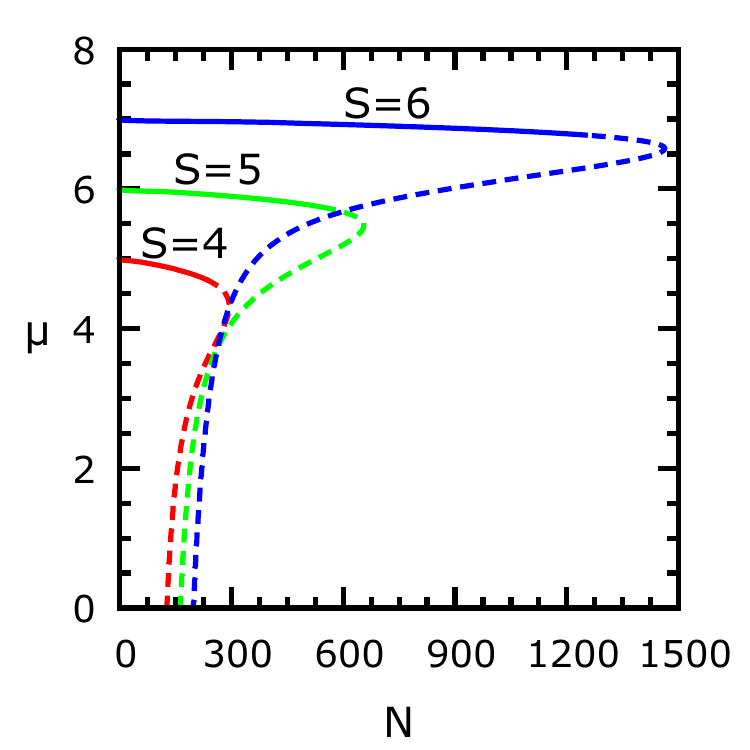}}
\caption{(Color online) $\protect\mu (N)$ curves for different vorticities $S
$, as produced by the numerical solution of Eq. (\protect\ref{gp_rad}) with $%
b=1.5$. Solid and dashed lines denote stable and unstable portions of the
trapped-vortex families.}
\label{mus}
\end{figure}

\begin{table}[h]
\begin{center}
Table 1: Intervals $\left( \mu _{\min };\mu _{\max }\right) $ in which the
vortices are stable, at different values of $S$ and $b$. For $S=1,3,4$, only
$\mu _{\min }$ is given, as $\mu _{\max }=1+S$ in these cases. \\[0pt]
\begin{tabular}{|c|c|c|c|c|c|c|}
\cline{2-7}
\multicolumn{1}{c|}{}&S=1&S=2&S=3&S=4&S=5&S=6 \\ \hline
b=0.75 & 1.21 & (2.16;2.27) & unst. & unst. & unst. & unst. \\ \hline
b=1.0 & 1.29 & (2.30;3) & unst. & 4.82 & (5.51;5.84) & (6.51;6.60)\\ \hline 
b=1.5 & 1.51 & (2.58;3) & 3.64 & 4.69 & (5.73;6) & (6.76;7) \\ \hline
b=2.0 & 1.66 & (2.73;3) & 3.79 & 4.83 & (5.85;6) & (6.84;6.96) \\ \hline
\end{tabular}
\end{center}
\end{table}

\begin{table}[h]
\begin{center}
Table 2: Intervals $\left( N_{\min };N_{\max }\right) $ in which the
vortices are stable, at different values of $S$ and $b$. For $S=1,3,4$, only
$N_{\max }$ is given, as $N_{\min }=0$ in these cases. 
\begin{tabular}{|c|c|c|c|c|c|c|}
\cline{2-7}
\multicolumn{1}{c|}{}&S=1&S=2&S=3&S=4&S=5&S=6 \\ \hline
b=0.75&11.0&(18;20.3)&unst.&unst.&unst&unst. \\ \hline
b=1.0&12.2&(0;25)&unst.& 27.4 & (42.5;103)&(150;172) \\ \hline
b=1.5 & 15.1 & (0;41) & 101 & 237 & (0;550) & (0;1274) \\ \hline
b=2.0 & 20.7 & (0;83) & 299 & 1066 & (0;3829) & (2160;13418) \\ \hline
\end{tabular}
\end{center}
\end{table}

Note that, for $S=5$ and $6$, Table 2 includes entries with the stability
intervals starting at finite $N_{\min }$, rather than at $N=0$. Accordingly,
in Table 1, these cases correspond to $\mu _{\max }<1+S$. This complex
behavior cannot be explained by simple arguments given above, and might only
be revealed by the systematic numerical analysis.

These results can be translated into physical units, taking the atomic BEC
of $^{7}$Li as an example, with $\Omega _{z}/2\pi \sim 1$ kHz, which
corresponds to $a_{z}\sim 3$ $\mathrm{\mu }$m, and the in-plane trapping
frequency $\sim 10$ Hz. Then, the actual number of atoms is estimated as $%
\sim 500N$ ($N$ are the values shown in Fig. \ref{mus} and Table 2), and the
size of the vortices, which are shown below in Figs. \ref{denstable}-\ref%
{denl4}, is obtained from the respective scaled lengths by multiplying them
by $\sim 30$ $\mathrm{\mu }$m. Thus, the critical values of the atom number,
corresponding to the data presented in Table 2, range from $\sim 5000$ for $%
S=1$ to $\sim 5\times 10^{7}$ for $S=6$. The latter estimate implies that
practically all the vortices with high values of $S$, which can be created
at relevant values of $b$, will be stable.

\textit{The linear-stability analysis.} The stability results reported above
were produced by taking perturbed solutions to Eq. (\ref{cl2deq}) as \cite%
{attractiveBEC}
\begin{gather}
\psi (r,t)=\left[ R(r)+\varepsilon \sum_{L}u_{L}(r)e^{iL\theta -i\omega
_{L}t}+\right.  \notag \\
\left. \varepsilon \sum_{L}v_{L}^{\ast }(r)e^{-iL\theta +i\omega _{L}^{\ast
}t}\right] e^{iS\theta -i\mu t},  \label{pertsol}
\end{gather}%
where $(u,v)$ and $\omega _{L}$ are eigenmodes and eigenfrequencies
corresponding to integer azimuthal index $L$ of the perturbation with
infinitesimal amplitude $\varepsilon $. The linearization around the
stationary solution leads to a system of the Bogoliubov de-Gennes equations
\cite{pethick},
\begin{equation}
\left( {%
\begin{array}{cc}
\hat{D}_{+} & gR^{2} \\
-gR^{2} & -\hat{D}_{-}%
\end{array}%
}\right) \left(
\begin{array}{c}
u_{L} \\
v_{L}%
\end{array}%
\right) =\omega _{L}\left(
\begin{array}{c}
u_{L} \\
v_{L}%
\end{array}%
\right) ,  \label{linearizedequ}
\end{equation}%
\begin{equation}
\hat{D}_{\pm }\equiv -\frac{1}{2}\left[ \frac{\partial ^{2}}{\partial r^{2}}+%
\frac{1}{r}\frac{\partial }{\partial r}-\frac{(S\pm L)^{2}}{r^{2}}\right] +%
\frac{1}{2}\Omega _{r}^{2}{r}^{2}+2gNR^{2}-\mu ,  \notag
\end{equation}%
supplemented by the boundary conditions requiring $u(r)$ and $v(r)$ to decay
as $r^{|S\pm L|}$ at $r\rightarrow 0$, and exponentially at $r\rightarrow
\infty $.

\begin{figure}[tbph]
\subfigure{\includegraphics[height = 53.0 mm, width = 53.0
mm]{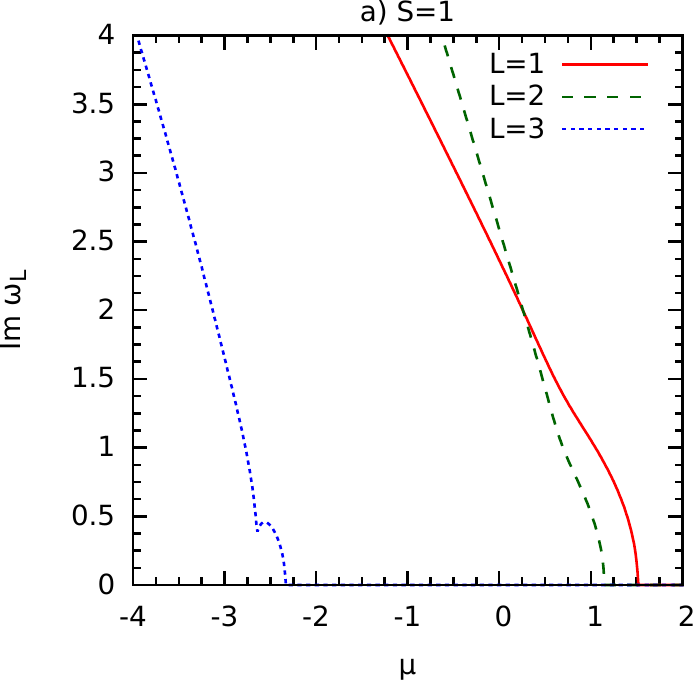}}
\subfigure{\includegraphics[height = 53.0 mm, width
= 53.0 mm]{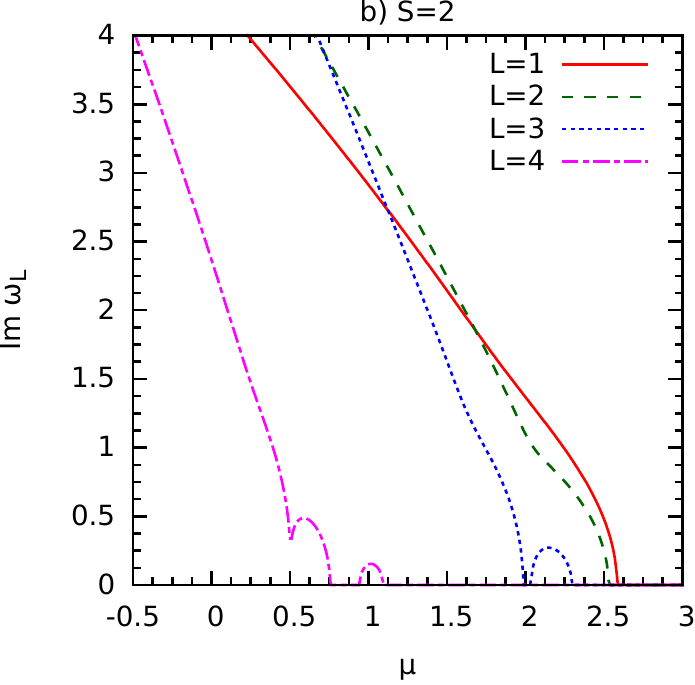}}
\subfigure{\includegraphics[height = 53.0 mm,
width = 53.0 mm]{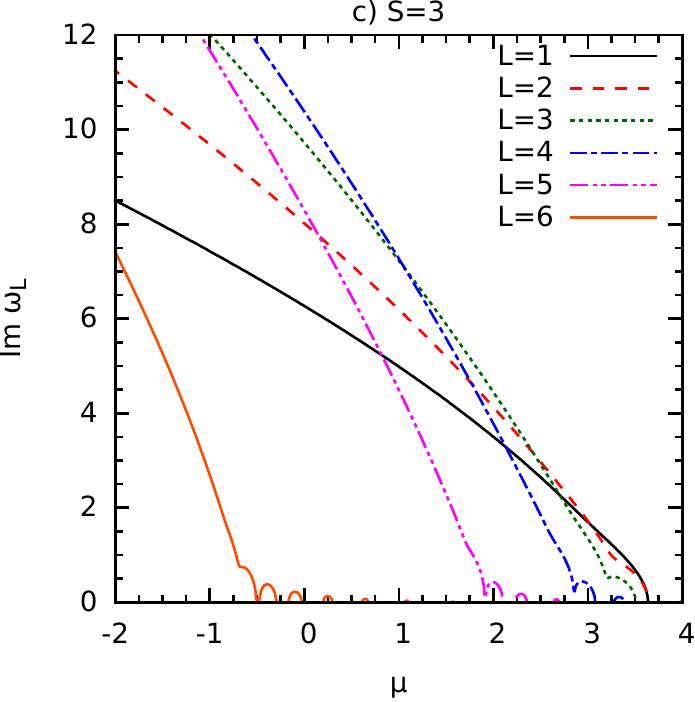}}
\subfigure{\includegraphics[height = 53.0 mm, width = 53.0
mm]{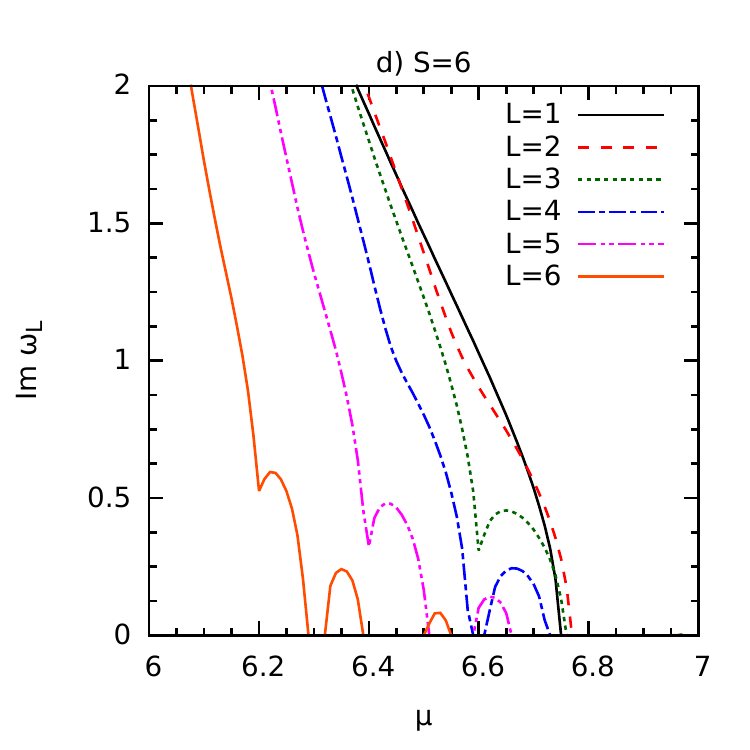}}
\caption{(Color online) The imaginary part of eigenfrequencies for different
values of azimuthal index $L$ of perturbation modes for vortices with $%
S=1,2,3$ and $6$.}
\label{egv}
\end{figure}

The determination of the stability from Eq. (\ref{linearizedequ}) requires
numerical diagonalization of the matrix. The vortex is unstable if, for
given $\mu $, at least one pair of eigenvalues $\omega _{L}$ is complex.
Results of the analysis are illustrated by Fig. \ref{egv}, which shows the
instability growth rates of perturbations, i.e., the largest imaginary part
of $\omega _{L}$, vs. $\mu $ for $S=1,2,3$ and $6$. We conclude that
stability regions exist for all these values of $S$, as shown above in
Fig. \ref{mus} and Tables 1 and 2.

\textit{Perturbed evolution of the trapped vortices}. To corroborate the
predictions of the linear-stability analysis, we have performed simulations
of perturbed evolution of the vortices in the framework of Eq. (\ref{cl2deq}%
), using the split-step and D'yakonov methods \cite{dyakonov,quinney}. We
have chiefly employed a grid with 400$\times $400 points and spatial spacing
$\Delta x=\Delta y=0.025$, the time step being $\Delta t=0.00005$. To ensure
robustness of the algorithm, the results were compared to those produced
with other values of $\Delta x=\Delta y$ and $\Delta t$. Initial random
perturbations with a $5\%$ relative amplitude were added in the simulations.

First, in Fig. \ref{denstable}, we display the perturbed evolution of a
stable vortex with $S=3$ and $\mu =3.8$ ($N=69$), in terms of the density
and phase. It exhibits rapid self-cleaning of the initially perturbed
configuration, which is observed in the stability regions for all values of $%
S$.

In the instability region, the simulations display fragmentation of
vortices, depending on azimuthal index $L$ of the respective leading
perturbation eigenmode, as shown in Fig. \ref{denl4}. In this figure, the
unstable vortex with $S=3$ breaks into four fragments, the corresponding
dominant eigenmode indeed having $L=4$. Eventually, the fragments suffer the
intrinsic collapse.

Thus, the direct simulations corroborate the predictions of the
linear-stability analysis for the trapped higher-order vorticities,
including the fact that the splitting of unstable vortices is driven by
dominant perturbation eigenmodes. It is worth to note that, unlike the model
with the uniform attractive nonlinearity \cite{attractiveBEC}, here the
stability and instability regions are not separated by a semi-stable one,
where the vortex with $S=1$ periodically splits and recombines, keeping the
vorticity intact.

\begin{center}
\begin{figure}[tbph]
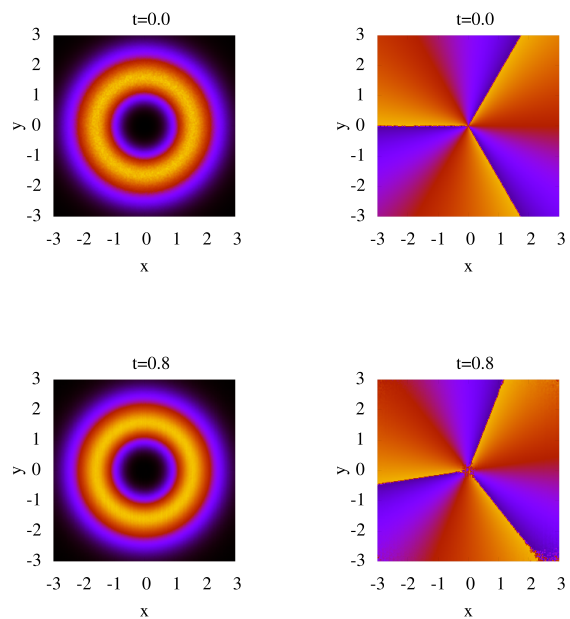

\subfigure{\includegraphics[height = 42.0 mm, width = 42.0
mm]{Fig_3_ds3est_1.pdf}}
\subfigure{\includegraphics[height = 42.0 mm, width = 42.0 mm]{Fig_3_fs3est_1.pdf}}
\subfigure{\includegraphics[height = 42.0 mm, width = 42.0 mm]{Fig_3_ds3est_2.pdf}}
\subfigure{\includegraphics[height = 42.0 mm, width = 42.0 mm]{Fig_3_fs3est_2.pdf}}
\caption{(Color online) The evolution of the density and phase for the
vortex with $S=3$ and $\protect\mu =3.8$ ($N=69$). Random perturbation,
added at $t=0$, is quickly eliminated by self-cleaning of the stable vortex,
becoming invisible at $t=0.8$, even if it is $\sim 0.1$ of the
characteristic diffraction time for the entire vortex. Here and below, in
density profiles, brighter regions correspond to higher density, and the
phase varies from $-\protect\pi $ to $\protect\pi $ from darker to brighter
regions.}
\label{denstable}
\end{figure}

\begin{figure}[tbph]
\subfigure{\includegraphics[height = 42.0 mm, width = 42.0
mm]{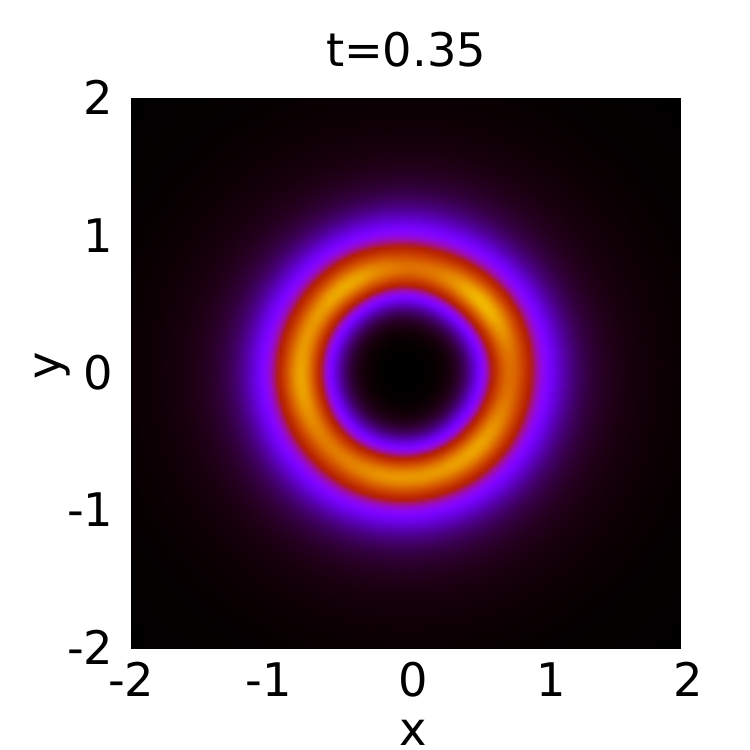}}
\subfigure{\includegraphics[height = 42.0 mm, width = 42.0
mm]{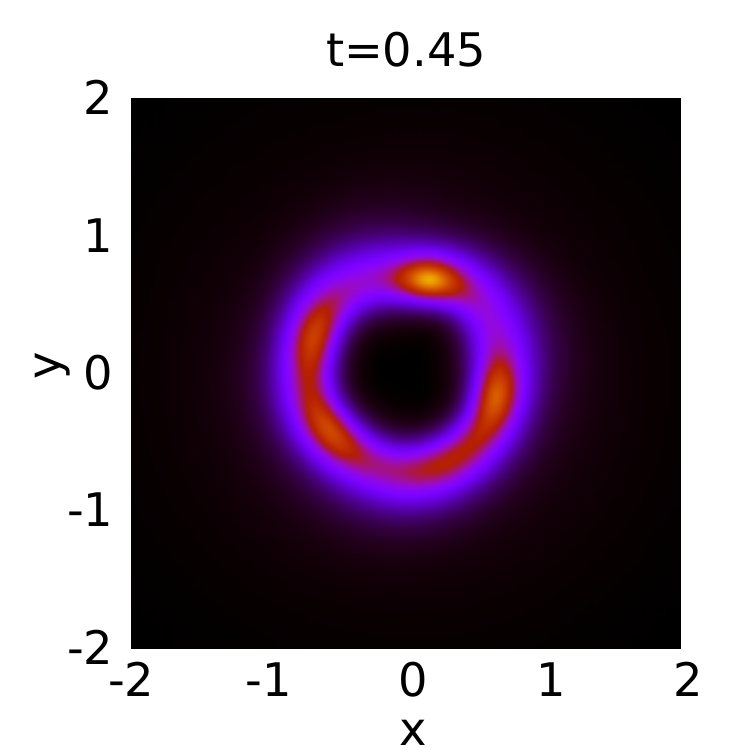}}
\caption{(Color online) Perturbed evolution of the density field for the
unstable vortex with $S=3$ and $\protect\mu =0$ ($N=90$).}
\label{denl4}
\end{figure}
\end{center}

\textit{Conclusions}. We have introduced the model of a 2D trapped BEC with
the self-attractive nonlinearity acting in a spatially localized manner.
Unlike the previously studied systems with the uniform self-attraction, in
which trapped vortices may be stable solely with topological charge $S=1$,
the present setting gives rise to stability areas for multiple vortices,
with $1\leq S\leq 6$. The largest radius of the nonlinearity localization,
which allows the stabilization of the higher-order vortices, was estimated
analytically, and accurately found in the numerical form, using the
linear-stability analysis and direct simulations. The number of atoms which
can be held by the stable vortex sharply increases with $S$. Challenging
problems for further analysis are the consideration of rotating multi-vortex
complexes (cf. Ref. \cite{Paderborn}) and generalization for the 3D setting.

\textit{Acknowledgments}. J.B.S. wishes to thank Council of Scientific and Industrial Research (CSIR),
India, for the financial support. R.R. acknowledges DST (grant No.
SR/S2/HEP-26/2012), Council of Scientific and Industrial Research (CSIR),
India (grant 03(1323)/14/EMR-II dated 03.11.2014) and Department of Atomic
Energy - National Board of Higher Mathematics (DAE-NBHM), India (grant
2/48(21)/2014 /NBHM(R.P.)/R \& D II/15451 ) for financial support in the
form of Major Research Projects. H.F. and A.G. thank FAPESP, CNPq and CAPES
(Brazil) for financial support.

\end{document}